\documentclass[aip,rsi,amssymb,amsmath,reprint]{revtex4-1}
\usepackage{color}

\usepackage{graphicx}
\usepackage[hidelinks]{hyperref}

\usepackage{orcidlink}

\usepackage[
    range-phrase=--,
    range-units=single,
    retain-explicit-plus=true,
    retain-unity-mantissa=false,
]{siunitx}

\newcommand{\CE}{Cosmic Explorer}
\DeclareSIUnit{\rtHz}{/\sqrt{Hz}}

\begin{document}



\setlength{\parskip}{1em}

\preprint{CE-P2400002}


\title[Cosmic Explorer Location Criteria]{Criteria for identifying and evaluating locations that could potentially host the Cosmic Explorer observatories} 

\author{Kathryne J. Daniel\orcidlink{0000-0003-2594-8052}}
\email[]{kjdaniel@arizona.edu}
\affiliation{Department of Astronomy \& Steward Observatory, University of Arizona, Tucson, AZ 85721, USA}

\author{Joshua R. Smith\orcidlink{0000-0003-0638-9670}}
\email[]{josmith@fullerton.edu}
\affiliation{The Nicholas and Lee Begovich Center for Gravitational-Wave Physics and Astronomy, California State University, Fullerton, 92831, USA}

\author{Stefan Ballmer\orcidlink{0000-0003-2306-523X}}
\affiliation{Department of Physics, Syracuse University, Syracuse, NY 13244, USA}

\author{Warren Bristol\orcidlink{0000-0003-2646-0315}}
\affiliation{Department of Geography, University of Arizona, Tucson, AZ 85721, USA}

\author{Jennifer C. Driggers\orcidlink{0000-0002-6134-7628}}
\affiliation{LIGO Hanford Observatory, Richland, WA 99352, USA}

\author{Anamaria Effler\orcidlink{0000-0001-8242-3944}}
\affiliation{LIGO Livingston Observatory, Livingston, LA 70754, USA}

\author{Matthew Evans\orcidlink{0000-0001-8459-4499}}
\affiliation{MIT Kavli Institute, Massachusetts Institute of Technology, Cambridge, MA 02139, USA}

\author{Joseph Hoover\orcidlink{0000-0003-2566-8042}}
\affiliation{Department of Environmental Science, University of Arizona, Tucson, AZ 85721, USA}

\author{Kevin Kuns\orcidlink{0000-0003-0630-3902}}
\affiliation{MIT Kavli Institute, Massachusetts Institute of Technology, Cambridge, MA 02139, USA}

\author{Michael Landry\orcidlink{0009-0000-0174-1247}}
\affiliation{LIGO Hanford Observatory, Richland, WA 99352, USA}

\author{Geoffrey Lovelace\orcidlink{0000-0002-7084-1070}}
\affiliation{The Nicholas and Lee Begovich Center for Gravitational-Wave Physics and Astronomy, California State University, Fullerton, 92831, USA}

\author{Chris Lukinbeal\orcidlink{0000-0003-1827-7764}}
\affiliation{Department of Geography, University of Arizona, Tucson, AZ 85721, USA}

\author{Vuk Mandic\orcidlink{0000-0001-6333-8621}}
\affiliation{School of Physics and Astronomy, University of Minnesota, Minneapolis, MN 55455, USA}

\author{Kiet Pham\orcidlink{0000-0002-7650-1034}}
\affiliation{School of Physics and Astronomy, University of Minnesota, Minneapolis, MN 55455, USA}

\author{Jocelyn Read\orcidlink{0000-0002-3923-1055}}
\affiliation{The Nicholas and Lee Begovich Center for Gravitational-Wave Physics and Astronomy, California State University, Fullerton, 92831, USA}

\author{Joshua B. Russell\orcidlink{https://orcid.org/0000-0003-3251-2919}}
\affiliation{Department of Earth and Environmental Sciences, Syracuse University, 
Syracuse, NY 13244, USA}

\author{François Schiettekatte\orcidlink{0000-0002-2112-9378}}
\affiliation{Département de physique, Université de Montréal, Montréal, Québec, Canada}

\author{Robert M. S. Schofield\orcidlink{0000-0002-6514-8693}}
\affiliation{Department of Physics, University of Oregon, Eugene, OR 97403, USA}
\affiliation{LIGO Hanford Observatory, Richland, WA 99352, USA}

\author{Christopher A. Scholz\orcidlink{0000-0002-7178-7619}}
\affiliation{Department of Earth and Environmental Sciences, Syracuse University, 
Syracuse, NY 13244, USA}

\author{David H. Shoemaker\orcidlink{0000-0002-4147-2560}}
\affiliation{MIT Kavli Institute, Massachusetts Institute of Technology, Cambridge, MA 02139, USA}

\author{Piper Sledge\orcidlink{0000-0001-9615-6912}}
\affiliation{Department of Gender \& Women's Studies, University of Arizona, Tucson, AZ 85721, USA}

\author{Amber Strunk\orcidlink{0000-0001-7707-4348}}
\affiliation{LIGO Hanford Observatory, Richland, WA 99352, USA}

\date{\today}


\begin{abstract}
Cosmic Explorer (CE) is a next-generation ground-based gravitational-wave observatory that is being designed in the 2020s and is envisioned to begin operations in the 2030s together with the Einstein Telescope in Europe. The \CE\ concept currently consists of two widely separated L-shaped observatories in the United States, one with \qty{40}{\km}-long arms and the other with \qty{20}{\km}-long arms. 
This order of magnitude increase in scale with respect to the LIGO-Virgo-KAGRA observatories will, together with technological improvements, deliver an order of magnitude greater astronomical reach, allowing access to gravitational waves from remnants of the first stars and opening a wide discovery aperture to the novel and unknown. In addition to pushing the reach of gravitational-wave astronomy, \CE\ endeavors to approach the lifecycle of large scientific facilities in a way that prioritizes mutually beneficial relationships with local and Indigenous communities. This article describes the (scientific, cost and access, and social) criteria that will be used to identify and evaluate locations that could potentially host the \CE\ observatories. 
\end{abstract}


\maketitle 


\section{Introduction}\label{sec:intro}

Gravitational waves --- predicted by Albert Einstein in 1916
--- are ripples in spacetime that travel at the speed of light and carry information about their astronomical sources. Observatories such as the National Science Foundation (NSF)-funded Laser Interferometer Gravitational-Wave Observatory (LIGO)~\cite{TheLIGOScientific:2014jea} and its international partners Virgo~\cite{TheVirgo:2014hva} and KAGRA~\cite{Akutsu:2018axf} have, through more than 90 observations of gravitational waves from black hole and neutron star mergers, opened the field of gravitational-wave astronomy~\cite{Abbott:2016blz,TheLIGOScientific:2017qsa,KAGRA:2021vkt}. Cosmic Explorer (CE) is the next-generation ground-based gravitational-wave observatory 
envisioned to begin operations in the US in the 2030s~\cite{evans2023cosmicexplorersubmissionnsf, CEHS}. \CE\ will use scaled-up technology based on the latest upgrades to LIGO (referred to as LIGO A+~\cite{KAGRA:2013rdx} and A\#~\cite{T2200287}) to deliver an order of magnitude greater astronomical reach, 
yielding scientific outcomes far beyond those of its predecessors~\cite{Corsi:2024vvr}.

The performance of any large physics experiment or ground based observatory and the success of its workforce depends on both environmental and social factors.
The criteria used to identify and evaluate locations that could host a \CE\ observatory must therefore be carefully considered.  
 
The \CE\ reference concept~\cite{evans2023cosmicexplorersubmissionnsf,CEHS} 
is an evolution of the LIGO design.  LIGO has two widely separated observatories based in Hanford, Washington  
and Livingston, Louisiana. 
A third LIGO observatory, LIGO India, is now under construction in Aundha in Maharashtra~\cite{Saleem_2022}.
Each of these hosts one laser interferometer with \qty{4}{\km}-long perpendicular arms (see Figure 1). Like LIGO, the \CE\ concept~\cite{CEHS} includes two widely-separated L-shaped observatories in the US. 
But the scale of \CE\ is much greater; it pairs one observatory with \qty{40}{\km} arms and far-reaching, broadband observations with another with \qty{20}{\km} arms to allow for gravitational-wave source localization and polarization information as well as tuning of its sensitivity to the (kilohertz) frequencies generated by neutron stars after they merge~\cite{evans2023cosmicexplorersubmissionnsf}.
\CE\ will achieve an order of magnitude increase in gravitational-wave amplitude sensitivity and a bandwidth that is widened toward lower frequencies with respect to the LIGO detectors. This level of performance, especially when operating in concert with the planned European next-generation observatory, Einstein Telescope~\cite{Punturo:2010zz}, 
will observe black holes and neutron stars across cosmic time, probe the nature of the most extreme matter in the universe, and explore questions on the nature of gravity and fundamental physics~\cite{Corsi:2024vvr,Maggiore:2019uih}.

The vision for operating the Cosmic Explorer observatories~\cite{CEHS} is also an evolution of the LIGO model.
This includes on-site maintenance and operation of the facilities and its systems and hardware as well as management, community partnerships, and the analysis that is required to make the data available to the broader scientific community and the public~\cite{evans2023cosmicexplorersubmissionnsf}.
To accomplish this, the observatories will be staffed by scientists, engineers, technicians, educators, and management. 
Gravitational waves arrive at all times and from all points in the sky and gravitational-wave detectors are omni-directional, i.e., they do not require dark skies or pointing to make observations.
For this reason, during the foreseen years-long \CE\ scientific observation periods, the observatories will be operated around the clock to enable observing, sharing data, and generating astronomical alerts.

The process \CE\ is adopting for identifying and evaluating promising locations for its observatories is informed by several historical precedents.   
The two United States locations for the LIGO observatories were chosen by the National Science Foundation (NSF) in 1992 following a years-long nationwide search and evaluation~\cite{Isaacson:2016,Fluharty:1992}. 
A NSF panel, led by former NSF Director John Slaughter, reviewed 171 potential configurations of widely separated pairs based on 19 candidate sites.  The review considered the characterization of each site and the criteria by which they were evaluated. In February 1992, NSF Director Walter Massey made the final selection~\cite{Isaacson:2016,Fluharty:1992}.
The LIGO India site was chosen by the Indian government in 
2021, also following a detailed location evaluation campaign~\cite{Saleem_2022}. 
As for the US LIGO observatories, it is expected that any final decisions of location for \CE\ will be made---taking into account the reports characterizing the locations---by the National Science Foundation and the broader United States government.

\begin{figure}
\begin{center}
\includegraphics[width=2.8in]{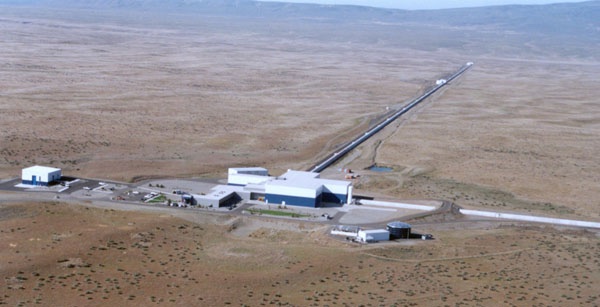}
\includegraphics[width=2.8in]{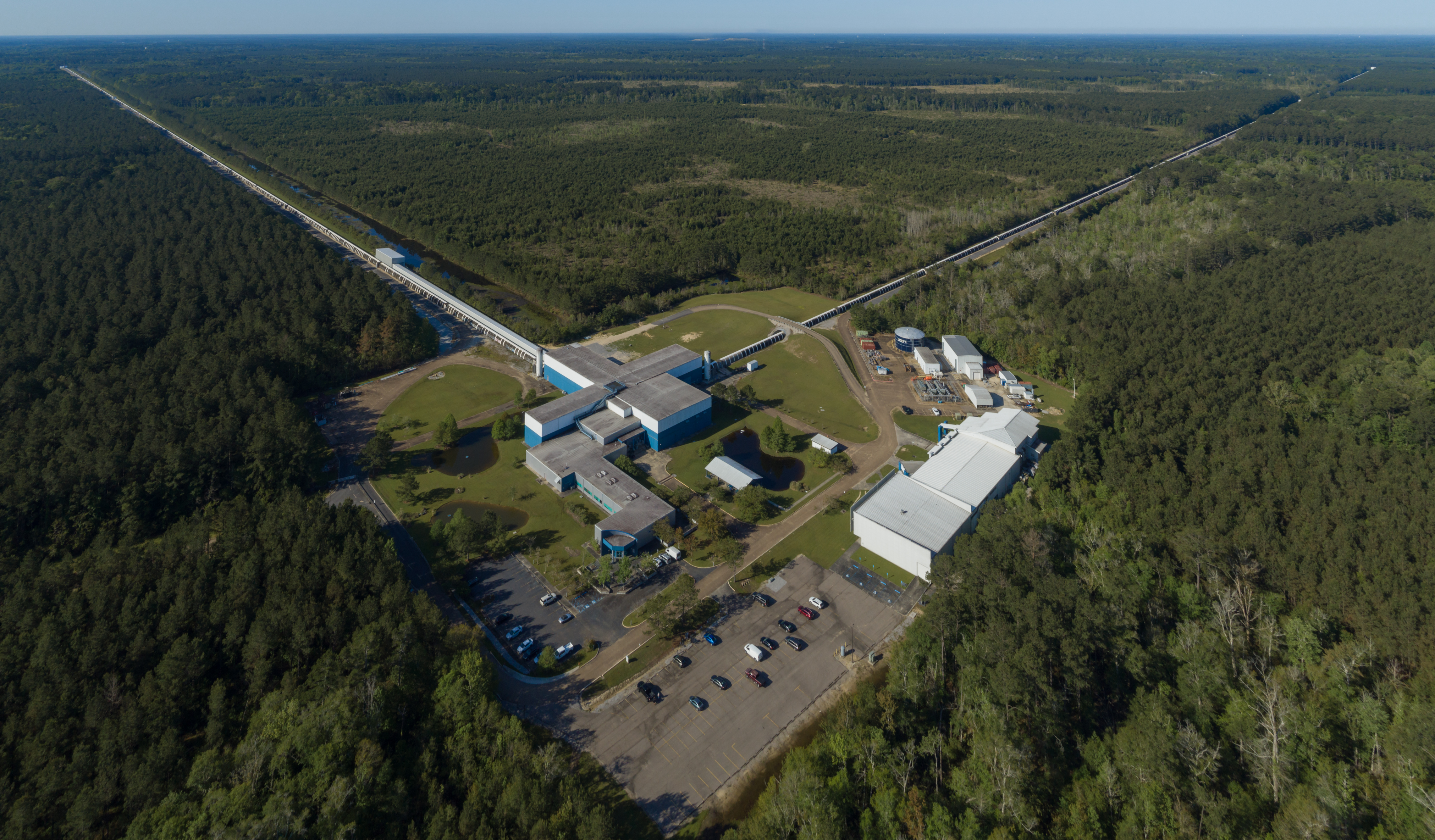}
\includegraphics[width=2.8in]{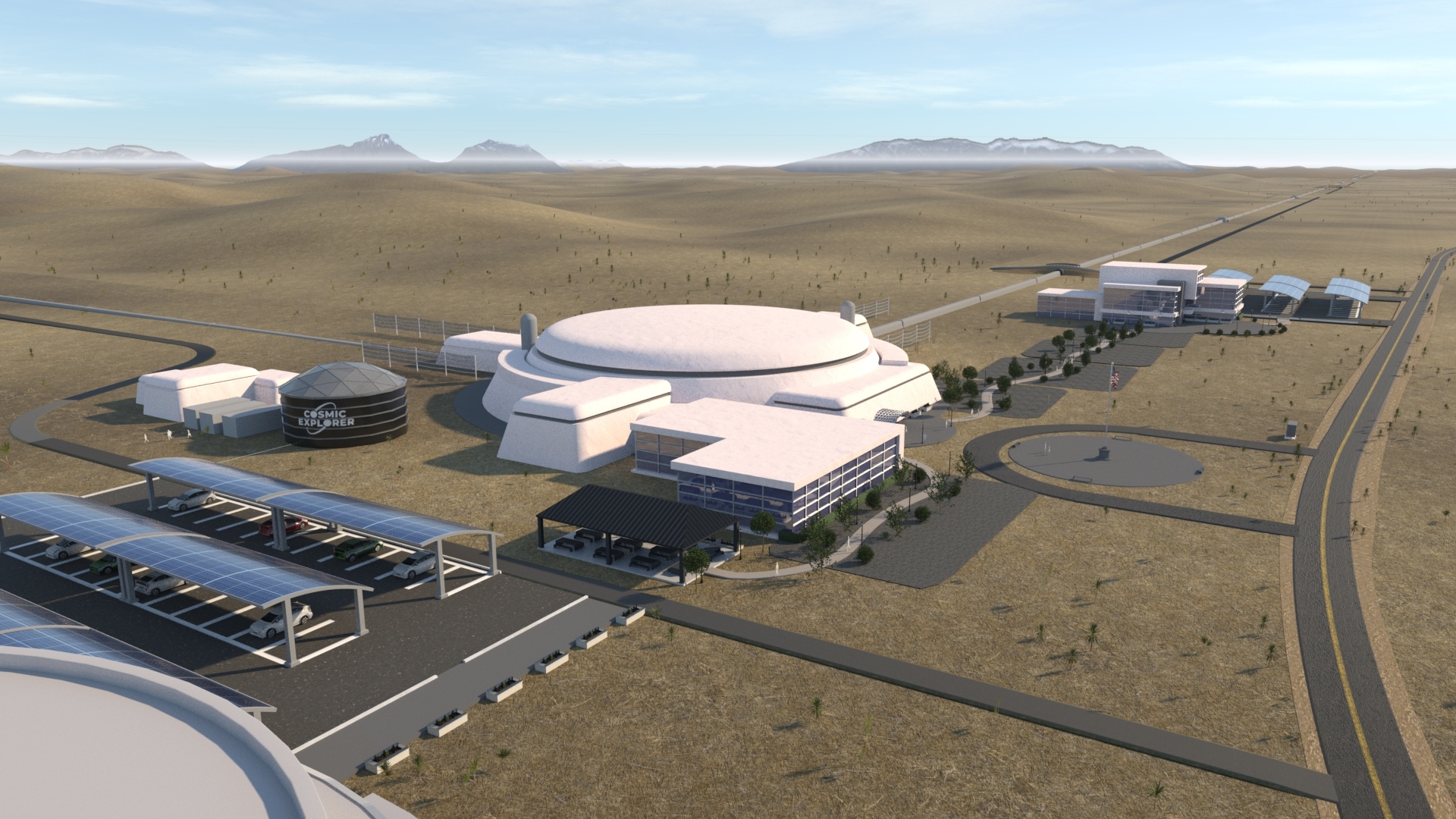}
\end{center}
\caption{Aerial views of the LIGO Hanford (top) and Livingston (center) observatories (credits: Caltech/MIT/LIGO Lab)\cite{TheLIGOScientific:2014jea} and at bottom an artist's impression of a Cosmic Explorer (CE) observatory (credits: Angela Nguyen, Virginia Kitchen, Eddie Anaya, California State University Fullerton)\cite{CEHS}.
}
\label{fig:observatories}
\end{figure}

The identification process for the Einstein Telescope is being undertaken in parallel, and lessons learned from that experience are informing \CE{}'s process.
As of 2024, Einstein Telescope's location criteria~\cite{10.1063/5.0018414} and identification process have produced candidate locations near the Euregio Meuse-Rhine~\cite{Koley_2022,Bader_2022}, an area near the borders of the Netherlands, Belgium and Germany, and Sardinia, Italy~\cite{Naticchioni:2024npy,Naticchioni_2020,Naticchioni_2014}. Scientific assessment of those locations is ongoing and an economic impact assessment has been performed for the Euregio Meuse-Rhine~\cite{ETimpact}.

Tensions between several major scientific facilities, such as TMT, Kitt Peak, and Mt. Graham, and their host communities have arisen surrounding issues of land acquisition, stewardship, and management.
These contentious relationships have resulted in negative associations with academic, scientific, and technical projects within those host communities~\cite{Nash2019Entangled, Lee2020LandGrab, Hodge2012No}, have led to significant risk for the facility success, and underscore the need for a revised approach (see recommendation from~\cite{Astro2020Decadal} and references therein).
Thus, \CE's approach to location identification and evaluation necessarily includes a thorough study of the historical and social landscape as well as early and consistent relationship building with local and Indigenous communities.
In recognition of this, new and upcoming facilities around the world have adopted a community based approach to siting and management \cite{binneman2021management, galaxies11010032, 2007EOSTr..88..309S}.

Identifying and evaluating locations for major scientific facilities is a multidisciplinary process that must take into account the potential scientific performance of the instruments, cost of construction and operations, and the environmental, cultural, and socio-economic landscape. 
A special focus is being placed on adopting criteria and a process used to identify and characterize promising locations for \CE\ that integrates conventional physical evaluation methods with a deeper understanding of the social, cultural, tribal, inter-tribal, and community landscape in which the observatories will be embedded.
The identification of potential sites, and subsequently the facility and workforce management 
will significantly invest in building long lasting, mutually respectful, culturally relevant relationships that center trust, process, and community interests \cite{david2018global, GVS,Astro2020Decadal}.  

This paper introduces criteria for identifying and evaluating locations for a \CE\ observatory and outlines an approach that integrates social, cultural, and physical principles from the outset. These criteria form the basis of a larger process for site identification that will be more fully described in later publications. A successful process will identify and characterize several locations where \CE\ can achieve its science goals (\ref{ss:science}), be built within appropriate cost boundaries (\ref{ss:cost}), attract, support and retain a diverse workforce, and where observatory activities can be aligned with community values (\ref{ss:social}). The resulting information will be shared with the NSF to facilitate site selection.


\section{Location Evaluation Criteria}\label{sec:factors}

The nature and scale of \CE\ means that the physical and socio-cultural elements of its surroundings are deeply intertwined. For this reason, the criteria that are used to identify and assess potential locations for \CE\ are herein presented together and will be considered together
from the start of the \CE\ location identification and evaluation process.  


\subsection{Science performance requirements}\label{ss:science}

To accomplish the science goals for \CE{}~\cite{evans2023cosmicexplorersubmissionnsf,CEHS}, promising host locations must each be able to accommodate an L-shaped observatory with equal arm lengths close to either \qty{40}{\km} or \qty{20}{\km} and the two final locations must be widely separated. There are also limiting environmental factors that constrain the potential performance of an observatory. 

\subsubsection{Observatory footprint}\label{ss:footprint}

The footprint of the \CE{} observatories is envisioned to be L-shaped with arms of length \qty{40}{\km} or \qty{20}{\km} and with a cleared width of \qty{75}{\m} along the arms to accommodate the beam tubes and their enclosures, small buildings (mid-stations and vacuum pumping stations), roads, and other civil engineering elements as needed such as land bridges for wildlife, over/underpasses, and drainage. Additionally, a roughly \qty{300}{\m} by \qty{300}{\m} campus area will be located near to the vertex of the arms (likely offset by some $\sim$\qty{1}{\km} distance to reduce human-induced noise coupling to the sensitive components at the vertex) to accommodate offices, education and community areas, parking, and staging areas.   

Preliminary geographical analyses indicate that allowing for small (few km) reductions in the length of the arms from the nominal \qty{40}{\km} and \qty{20}{\km} values and allowing other opening angles each significantly increase the number of locations that can accommodate \CE{}'s footprint with acceptable cost while retaining most of the target sensitivity (as described in Sec. \ref{SecCivil}).  The impact of changes to the detector length and angle on \CE's performance is complicated and depends on both the detector configuration and on the sources being observed. It also depends on the characteristics of the network the detector is augmenting, as the science reach of the network is determined by the location and sensitivity of all of the network nodes (whose signal to noise ratios combine in quadrature).

Following~\citet{Read2023}, we consider the impact of a new detector on representative science metrics including the volume explored, number of sources, and signal to noise ratio of the loudest events, compared with a detector of the same noise shape but different length and angle configuration. The sensitivity of an interferometer to a given gravitational-wave signal can be approximated as proportional to the length of the arms $L$ and to the sine of the opening angle of the arms $\theta$, so a \qty{90}{\degree} opening angle maximizes the signal for a given arm length. When a \qty{40} or \qty{20} {\km} facility is added to a network of 1-2 other comparable scale facilities, the science metrics we consider scale as $\left( L_{\rm{new}} \sin \theta \right)^{\alpha}$ where the coefficient $\alpha$ is between $1.4$ and $2.5$. This means that locations that require length reductions or opening angle deviations from \qty{90}{\degree} will be somewhat disfavored. As rules of thumb, a length reduction of more than \qty{2}{\km} or opening angles outside the range \qtyrange{65}{115}{\degree}, would each give more than a \qty{10}{\%} reduction of sensitivity. 

The \CE{} optics will be isolated  from ground motion with actively stabilized platforms and via pendulum suspensions. While the gravitational wave readout is primarily sensitive to the longitudinal motion of the optics, vertical displacement (caused by seismic and thermally-driven motion of the suspensions) couples into the readout if the optics hang at an angle to the main laser beam. To minimize this coupling, the net tilt of a potential site, measured from end-to-end of the \qty{40}{\km} or \qty{20}{\km} arm, respectively, is targeted to be less than \qty{3}{\m/\km} and \qty{1.5}{\m/\km}, which corresponds to the tilt that the pendulums of a perfectly flat detector would have from hanging towards the center of the earth, $\phi_0 = L_{\rm{arm}}/2R_{\rm{earth}}$~\cite{Kuns2020}.

The scale of a \CE\ observatory makes it imperative that locations be found which minimize the disturbance to the land, since reshaping the land drives costs for observatory construction and has myriad cultural and environmental ramifications. To start to identify appropriate sites, we will initially use public elevation and geological data and the parametric cost model developed for the \CE{} Horizon Study \cite{CEHS} to estimate the civil engineering costs of any given observatory footprint, with the understanding that other factors will require additional information and more in-depth consideration. In particular, as relationships develop around a given candidate location, all the above will be discussed with local and Indigenous communities in order to co-produce community knowledge and understanding of the environment and cultural considerations.
For example, adjustments to the facility length or opening angle may be required, or a candidate location may simply be ruled out, to avoid disturbing culturally or environmentally sensitive areas.

\subsubsection{Global separation and orientation}
The reference concept for \CE{} calls for two widely separated observatories, one with \qty{40}{\km} arms and one with \qty{20}{\km} arms~\cite{CEHS}.
The scientific output in general, and sky localization for gravitational-wave events in particular, depend on the separation and relative orientation between the \CE{} observatories. For example, the polarization properties of gravitational waves are such that a maximum of information would be obtained with detectors close to 45 degrees to each other. Similarly, for purposes of localizing the direction of a gravitational wave source on the sky, as well as to minimize correlated environmental noise effects in detectors, it is necessary to keep detectors geographically separated. As an example, LIGO Livingston and Hanford Observatories are separated by \qty{3000}{\km} (\qty{10}{\ms} light travel time) and are rotated roughly \qty{90}{\deg} with respect to one another. This choice of 90 deg was imposed to ensure that first detections could be internally verified with the two LIGO detectors, but following LIGO-Virgo's observational success, the more scientifically rich choice of 45 deg is preferred for future detectors.

It is important to note, however, that CE observatories will likely operate as a part of a broader network of gravitational wave detectors, which could include LIGO-India, Einstein Telescope, and other next-generation facilities. The number and the sensitivities of these detectors at the time of CE operation are currently uncertain, but it is clear that their presence will provide additional information (polarization, timing/directionality) about gravitational waves, which will correspondingly alter the optimal positioning and orientation of the CE detectors and further boost the network's science output.

A recent report by the NSF Mathematical and Physical Sciences Advisory Committee Subcommittee on Next-Generation Gravitational-Wave Detector Concepts~\cite{Kalogera_Report} has investigated different detector network configurations involving \CE\ in the context of their science potential. The report considered multiple science objectives including observations of binaries of black hole and/or neutron stars across the universe, probing the dynamics of dense matter in neutron stars, multi-messenger observations and their applications to measure the Hubble constant and probe the dark sector of the universe, and others. In absence of the Einstein Telescope detector, the report demonstrates the need to have \emph{both} CE detectors in the network supplemented by a third detector geographically distant from the others (such as LIGO-India) with the sensitivity of Advanced LIGO upgrades (A+ or A\# sensitivity~\cite{T2200287}). If Einstein Telescope is a part of the detector network, then a single CE detector with \qty{40}{\km} arms is sufficient to reach all of the science objectives, if combined with a third detector at A+ or A\# sensitivity. Consequently, the CE locations, arm lengths, and arm orientations will be chosen so as to optimize the science output of a broader network of gravitational wave detectors.

\subsubsection{Environment and geology}

Components of the environment, including ground-cover and geology, affect the location suitability, construction costs and potentially the operations of \CE. Favorable locations require fewer changes to the landscape, which will reduce the overall civil engineering costs and may also minimize community concerns around such operations. Publicly available elevation and landcover data can be used to identify locations of interest that would have minimal impact on the land, and geological and geophysical data provide insight into the local environment.

A number of features must be considered to determine the long-term suitability of potential observatory locations.
This includes investigating proposed future developments around the location (e.g., mining, wind farms, or industrial or urban encroachment) that might alter the environment as well as the impact of climate change over the 50-year lifetime of the observatories. The risk associated with catastrophic natural disasters (floods, fires, earthquakes, etc.) will be estimated from publicly available data such as the National Risk Index published by the Federal Emergency Management Agency, which includes 18 natural hazards and quantifies expected annual loss while accounting for social vulnerability and community resilience at the county level. The U.S. Climate Vulnerability Index provides another metric for assessing long-term viability of a location and risks due to climate change and includes both physical and socioeconomic factors \cite{TeeLewis2023}. Indeed, climate resilience and climate sustainability will be key design considerations for \CE, taken up by the broader project design.

Seismicity and seismic noise are perhaps the most crucial environmental factors that affect nominal detector performance and its long-term stability. The United States Geological Survey (USGS) National Seismic Hazard Model \cite{Peterson_2023} provides physics-based probabilistic estimates of peak ground shaking expected at a given location within a time-span of 50 to 100 years. This information will be useful for choosing a location with low rates of seismicity and in the engineering stage to ensure earthquake resilience of the facility. In addition, existing high-quality 3-component broadband seismic data are openly available that span the United States from geophysical experiments such as the Earthscope USArray~\cite{Anthony:2022} and others. The USArray provides information about noise conditions resulting from both natural and anthropogenic causes at a spatial sampling of $\sim$70~km, and spanning the entire country.

The local geology and subsurface structure can greatly impact engineering costs as well as seismic noise observed at Earth's surface. Soft, thick sediments and sedimentary rocks can lead to ground motion amplification and resonance at distinct frequencies that depend on the elastic properties of the material and mode content of the seismic wavefield. On the other hand, while solid bedrock provides a more stable platform for construction, distant seismic waves can propagate more efficiently through it, negatively impacting detector performance. Both factors will be taken into account as preliminary analyses of the subsurface properties and geology are carried out. More detailed geotechnical assessment (including soil and subsurface conditions, hydrology and drainage) is ultimately needed for accurate construction cost and contingency estimates, along with well identified risks and potential mitigation strategies.
 
Locations that will allow \CE\ to be built and operated with as little negative impact on the land and environment as possible are preferable. Any candidate \CE\ site must undergo an environmental impact assessment following the National Environmental Policy Act (NEPA) and a process following The National Historic Preservation Act (NHPA) to assess and protect historical and cultural resources.
An observatory's carbon footprint will have some dependency on its location and thus sites that can lower or offset carbon emission are favored. 
Options for sharing space with an energy producer, such as a solar farm, will also be considered. 
However, energy producers such as wind turbines or hydroelectric tend to produce significant seismic and acoustic noise.

An area of focus not always captured by environmental and cultural impact assessments are the cultural impacts of changes to the environment due to facilities construction. 
The Indigenous and Place-based Partnerships \& Responsible Siting team (IPP-RS, \S\ref{s:IPP}) will focus on building a foundation of trust with Indigenous communities that will enable open knowledge exchange. By fostering these relationships from the start, we can understand the local and Indigenous communities' connections to the landscape and incorporate those perspectives and knowledge in the evaluation of \CE's cultural impacts. This includes potentially considering a location unsuitable for \CE\ if deemed at odds with the local communities' interests and/or would cause a negative cultural impact.

\subsubsection{Environmental limits to sensitivity and up-time}

The sensitivity and up-time of gravitational-wave observatories are influenced greatly by the environment in which they exist~\cite{Fiori:2021gsz,AdvLIGO:2021oxw,Virgo:2022ypn}.
The infinitesimal displacements that need to be measured to detect gravitational waves result
 in strong environmental sensitivity despite the interferometer's vacuum enclosure, and its elaborate isolation
 systems.
Ground vibrations can couple into the detectors mechanically (``seismic noise'') and, along with atmospheric disturbances, gravitationally (``local gravity noise''~\cite{harms2019terrestrial,Driggers:2011aa}).
Wind and rain can produce acoustic and seismic noise.
Electromagnetic noise, and radio frequency interference, can couple to the readout and control systems, and directly to the interferometer optics.
Gravitational waves arrive at earth at all times and the up-time of the detectors, which is determined by the fraction of each day that the interferometers are locked (i.e., operating nominally and sensitively with light resonating in their arms and servo control loops for their many length, angular, and other degrees of freedom closed), can be degraded by environmental disturbances that cause the servo loops to exceed their limits.  

Seismic noise of the locations must be low enough for \CE\ to achieve its target sensitivity and up-time.
The current estimate of the \CE\ sensitivity assumes a Rayleigh-wave-dominated ground motion with amplitude \qty{1}{(\um/\s^2)\rtHz}~\cite{CEHS,Hall:2020dps}.
Seismic noise in the 1-30\,Hz range has the potential to directly couple into the instruments through the seismic isolation systems, or indirectly through control signals, and stray light. Additionally, low frequency seismic noise down to \qty{10}{\mHz} can be highly variable, and can lead to up-converted noise in the detectors especially through the motion induced between the very distant buildings. 
The ``microseismic peak'' from \qtyrange{0.1}{0.3}{\Hz} can cause issues with interferometer control when its amplitude is around or above \qty{1}{\mu\m}.
Motion from the earth tides, while large, can be accurately estimated and counteracted in the instrument control systems.
Locations that have generally low overall ground motion amplitude from \qty{10}{\mHz} to \qty{30}{\Hz}, for example close to the Peterson Low Noise Model~\cite{peterson1993observations}, are preferable. National data exist in the US that will allow an assessment of this noise and its variability~\cite{Anthony:2022}.

Seismic noise and atmospheric pressure fluctuations (infrasound) must be low enough to not limit \CE{}'s sensitivity through local gravity noise (after subtraction)~\cite{harms2019terrestrial, Driggers:2011aa}.  
At LIGO Hanford, winds of more than \qty{10}{m/s} have been associated with decreased up-time and sensitivity~\cite{Ross:2020bqv}. While \CE\ may incorporate mitigation strategies such as wind fences and rounded buildings, locations with winds that rarely exceed that value are preferable. 
Potential observatory sites should also be free of strong electromagnetic disturbances at audio frequencies (which can couple directly into \CE{}'s sensitive band) and at radio frequencies (which can interfere with modulation and demodulation techniques for \CE{}'s interferometer control).

While existing sources of environmental data will be useful, they may miss certain performance-degrading environmental signals, and will need to be supplemented with local investigations. For example, existing seismic data sources may not capture local noise sources near a site (such as road traffic and off-road recreational vehicle traffic) and may not account for common but short transients that can cause problems for the detectors. For example, a couple of large trucks passing near a site per hour may significantly degrade the performance of the detector and yet, because the large signals are present for only tens of seconds per hour, they may be in the top one percentile of the seismic data. As a result, these important signals may not be present in summaries that include up to the 95th percentile.  In summary, both existing long term environmental data and focused data from site investigations are essential.


\subsection{Cost boundaries and access}\label{ss:cost}

The choice in location for the \CE\ sites will play a significant role in the project's overall cost due to civil engineering costs, the availability of surrounding infrastructure, and land acquisition. Access to the locations will follow an iterative process with the local communities and it will be important to obtain information on land rights and ownership, obtain permits for physical access, and to be cognizant of the interests, contexts, and protections of Indigenous people. 
Like LIGO, \CE\ is foreseen to be a project that benefits greatly from contributions by a vibrant international scientific community. Thus access by foreign nationals to the eventual \CE\ sites throughout the project lifecycle is required.

\subsubsection{Civil engineering cost factors}\label{SecCivil}

The majority of the civil engineering costs associated with building \CE{} at a given location will be driven by the physical characteristics of the site, especially in providing straight and level long arms
for the roughly \qty{1}{\m}-diameter vacuum pipes that will enclose the laser beams. 
Given the expected remoteness of promising locations, there will be costs associated with providing basic infrastructure including power, water, waste treatment, and high-speed internet. 
Additionally, there may be location-dependent costs associated with materials and construction and maintenance of the roads, buildings, and other infrastructure and safety and security provisions for the location.  
The electrical power usage of one \CE\ observatory, not including scientific computing clusters, is expected to be similar to the power used by the LIGO observatories, roughly one megawatt 
average continuous consumption.
The basic parameters of the power system may lead to location-dependent capital and long-term operation costs.
Suitable locations must present a combination of these location-dependent cost factors that are within the financial boundaries set by anticipated funding. 

\paragraph{Locations that reduce construction costs and changes to the land}
When searching for a location to install a pair of tubes in which a laser beam will propagate over a distance $L=40$\,km, probably the most significant civil engineering cost is to achieve a straight and level path for the vacuum system. For such lengths, the curvature of the Earth comes into play. If  a \qty{40}{\km} tube were installed at a location with constant elevation (with respect to the sea level), because of Earth's radius $R$, one  would have to dig a trench reaching $L^2/8R=$ 31\,m in depth (103\,ft) in the middle to achieve a straight line between the ends. Assuming a platform $w=4$\,m wide at the bottom of the trench and 45° walls on each side, to a good approximation, the digging volume for one arm is given by
\begin{equation}
    V = \left( \frac{1}{15} \left(\frac{L}{R} \right)^2 + \frac{2}{3}  \frac{w}{R}\right) \left(\frac{L}{2}\right)^3,
\end{equation}
which is nearly 50 million m$^3$ of earth to displace for both 40\,km arms, at a cost of several hundred million dollars, before considering the infrastructure to drain such a trench. We note that this expression indicates a digging volume that scales as $L^5$. As a consequence, a 10\% change in arm length represents a $\sim$40\% change in digging volume, on land with constant elevation. 
Alternatively, building trenches 11\,m deep in the middle and using that earth to build berms that would reach 20\,m high near each ends would reduce the displaced earth volume to about 11 million m$^3$. 

The volume of earth that would need to be displaced can  be significantly reduced by finding gentle valleys (in elevation) where the surface is actually flat in the Euclidean sense. 
Preliminary studies based on elevation maps suggest that many locations exist in North America where the total volume that would need to be displaced is as low as 1 million m$^3$ and that these locations tend to be such that they could accommodate several slightly different detector placements and orientations which means that there should be flexibility at a given location to avoid obstacles or accommodate other local considerations. Such locations, which may be associated with lower construction costs and require fewer changes to the land, are preferable. 

Additionally, the type of landcover that would overlap the footprint of the detector will be important for reducing costs and changes to the land. For example, construction across major highways, railways, bodies of water, forests, and wetlands would have civil engineering and/or environmental costs that would need to be taken into account.   

\subsubsection{Surrounding infrastructure}

The potential for a location to support a \CE{} observatory depends on access to a diverse collection of infrastructures.
Viable observatory locations must be secure and reasonably accessible via roads and airports and must not be frequently rendered inaccessible, for example by weather.
Nearby transportation infrastructure (highways, railroads, etc.) is advantageous during the construction phase, but may be disadvantageous during operation due to the resulting seismic noise. 
Similarly, \CE\ staff and visitors will require access to social infrastructure (schools, hospitals, universities), while the observatories require some distance from large-scale human activities to avoid the resulting environmental disturbances.
Experience with the LIGO observatories has shown that \qtyrange{15}{30}{} miles (\qtyrange{24}{48}{\km})  is a workable distance from population centers to gravitational-wave observatories.
Given the larger scale of \CE\ it is important, when considering this distance, to specify that the sensitive scientific equipment will be located at the vertex of the arms and at the end of the arms, while the office, and visitor/educational infrastructure may be located close to the vertex but distant enough (\qtyrange{0.2}{1}{\km}) to not cause significant disturbances.

\subsubsection{Land rights and permitting}

Identifying accessible lands for hosting the \CE\ observatories will bring a unique set of challenges. 
As described in previous sections, locations which are physically favorable for \CE\ require long, uninterrupted, relatively flat terrain along the L-shape that would be traveled by the laser beams along the observatory arms, and as such they tend to be either very remote, already in use (as national parks, military facilities, etc.), or both.
In addition to physical feasibility, eventual sites will require the support of the community.
It will thus be necessary to compile a comprehensive study of the physical, social, community, and legal landscapes for potential locations.  
These data will include relationships between various entities, if there are contested land permits and titles, whether there are different surface and subsurface titleholders, and other access considerations such as security clearance or citizenship requirements.

\subsubsection{Land acquisition}

Land acquisition costs are expected to play an important role in the overall cost of establishing the \CE\ sites. Land ownership along the footprint of a \CE\ observatory may be varied and complicated, ranging from, e.g., fully federally-owned land, to mixture of federal, state, Bureau of Land Management, tribal, and private lands.
Particularly the western US states which follow the Public Land Survey System (PLSS), also known as Township and Range, often have a variety of land ownership. For acquiring land, ownership, lease, or other models are possible. While costs may be roughly estimated at the national level, we expect land acquisition costing to be highly localized.  On the local level cost projections can be made in anticipation that they will come into focus at later stages of the project.

\subsubsection{Indigenous Peoples protections, interests and context}\label{s:FPIC}

All lands within the United States are the ancestral homelands of Indigenous Peoples~\cite{DunbarOrtiz}. As a reflection of this, access to and evaluation of locations that could potentially host \CE\ must be carried out in consultation with Indigenous Peoples and with respect for their protections, interests, and context. This is not as much a criterion of a given location as it is a consideration to be applied to all locations. 

As described later in Section~\ref{sec:approach}, the broader process of evaluating locations will be iteratively carried out by \CE\ personnel in conjunction with Indigenous and local communities, respecting the rights of those communities, and seeking out and considering their points of view. 
This process is structured to be in accordance with the United Nations guidance on Free, Prior, and Informed Consent (FPIC), the United Nations Declaration on the Rights of Indigenous People (UN-DRIP), and the American Declaration on the Rights of Indigenous Peoples (ADRIP), and to promote the long-term success and community synergies of the \CE\ observatories. 


\subsection{Social factors}\label{ss:social}

Site evaluation for the \CE\ project seeks to integrate conventional physical evaluation methods with a deeper understanding of the social, cultural, tribal, inter-tribal, and inter-community landscape in which the observatories will be embedded. The success of \CE\ will rely on attracting, supporting, and developing a diverse workforce that is in part from and also integrated with the local communities. This section describes some of the factors that will help determine a given location's suitability in these regards. 

\subsubsection{Communities and their interests}

We will analyze a number of socio-cultural characteristics obtained through public data.
Compiling these data will provide a first glimpse into the demographic and ethnographic traits of a given area and the first-hand experiences of the people. This approach also allows \CE{} to consider the location specific context and to illuminate opportunities for relationship building between Indigenous, local, and scientific communities. 
In other words, this will allow us to explore how \CE{} can align with the values of people from a particular location who will have direct experiences that are rooted in those specific lands and communities and their histories and ecosystems. 
The intention behind this approach is to identify the potential for long lasting relationships that support practices that are culturally compatible with local hosts. 

Our approach to evaluation of each site aims to identify major elements of the social landscape and community perspectives.
Understanding the history of a given location, including of its past, current, and future residents, will allow for better assessment of possible synergies between \CE{} and host communities. 
Considerations include removed and displaced Indigenous communities who retain ties to and interest in the land, minoritized communities, and debates regarding land use. 
Educational and scientific opportunities are a focus for \CE{} and can include partnerships with K-12, tribal colleges and universities (TCUs), minority serving institutions (MSIs), and research institutions.  These partnerships are an investment in both the \CE{} science and technical workforce as well as the future generations of local communities.
The significance of the land must be considered, including current and historical land management, uses, cultural connections, and environmental context.
The economic profile of locations will be important to evaluate potential two-way impacts between \CE{} and the community, such as job creation and engagement of local expertise.
We will also consider military, government, and legal landscapes in context of that place. 
These criteria can be brought together to find areas that have the potential be be both physically suitable for \CE{} and socially positive along multiple axes for the lifetime of the project.

\subsubsection{Quality of life}

\CE\ will be a hub for a diverse, international, and local workforce, as well as their families, and for visitors. Attracting, retaining, and supporting this workforce and community requires attention to factors such as housing markets, food access, quality educational and career opportunities for the families of CE staff, access to medical care, community resilience, safety and security. To gain holistic insight into potential sites, \CE\ will consider national and regional indices---compiled by federal agencies---that combine many such considerations to give general impressions about quality of life.

These national data will help clarify additional questions we may have about a community and may suggest opportunities for CE to positively support local and Indigenous communities. Because quality of life is subjective, these indices are not intended to exclude areas from participation in \CE\ but to guide researchers to areas where the project can develop positive local synergies  
and strengthen workforce outcomes.

\subsubsection{Social landscape}\label{s:SocialLandscape}

Social landscape considerations for \CE\ are intended to identify regions where the \CE\ workforce will be most supported while ensuring local and Indigenous communities are respected and included. These criteria follow from the Astro2020 Decadal Survey~\cite{Astro2020Decadal}, NSF Broader Impacts~\cite{NSFbroaderImpacts}, NSF Strategic Plan~\cite{NSFstrategicPlan2022}, and UN Social Development Indicators~\cite{UNSDGindicators} highlighting the scientific community’s awareness that large facilities must engage communities as collaborators and co-visionaries throughout the life cycle of those facilities. \CE’s process utilizes these and other criteria to inform an innovative approach to community relations while situating CE in the context of the development needs and goals of a community, the scientific goals of the facility, and the need to provide a pathway to creating and supporting a diverse and talented scientific workforce.

In collaboration with the National Opinion Research Center~\cite{NORC}, \CE\ is developing indicators for understanding community perspectives. 
These will illuminate how those communities view science and scientific facilities and will illustrate social inclusion/exclusion along multiple axes, social participation, social cohesion, institutional inclusion, and tolerance. 
An understanding of the regional social landscape will also provide context through which \CE\ can develop a preliminary view on the scope of opportunities and obstacles to building relationships with, and becoming part of, the relevant communities. 

\subsubsection{Indigenous and Place-based Partnerships \& Responsible Siting (IPP-RS)}\label{s:IPP}

Building, strengthening and maintaining positive, mutually beneficial relationships with Indigenous communities is a necessary criteria for successfully siting \CE{} and a central aim of for \CE's Indigenous and Place-based Partnerships \& Responsible Siting (IPP-RS) team.
It is from these relationships that trust can be established and upheld, thus providing a foundation from which mutually beneficial legal agreements, collaborative grant proposals, and a holistic approach to facilities operations can be established and maintained \cite{GVS,Venkatesan2019}. 
\CE's effort will go beyond compliance with institutional, local, state, federal and international regulations and protocols, thus ensuring accountability throughout the lifetime of the observatories.
These partnerships will be based on ongoing integration of and collaboration with communities in order to build a foundation for how \CE\ engages with communities and regards the land.
These partnerships will support mutual stewardship efforts and the specific goals outlined in Appendix N of the Astro2020 Decadal Survey~\cite{Astro2020Decadal}. Community representatives and members may include people from Native Nations, Indigenous communities and their governmental and cultural leadership, local community groups, local governmental jurisdictions, educational institutions, and any other community with an investment in the location of interest.

IPP-RS addresses an important and specific need of the \CE\ project but its potential reaches further. Globally, there is increasing, over-due recognition that the rights of Indigenous peoples are inalienable, and, furthermore, that Indigenous science has unique contributions to the many problems facing humanity (e.g.,~\cite{cajete2000native,norgaard2019salmon,atalay2012community,hernandez2020weaving,UNDRIP,CSIROindig,2022Natur.606..447C}). 
Lessons learned, partnerships made, and knowledge created through the collaboration between \CE\ and Indigenous communities will be documented throughout, adhering to the principles of Indigenous data sovereignty and governance \cite{UNDRIP}.
These outcomes could inform processes for other large-scale scientific facilities, provide example frameworks that Indigenous communities could build from to approach the scientific community for collaborative projects, and more broadly support positive relationships between Indigenous and STEM communities.
Regardless of whether a potential location is selected as an observatory site, \CE's approach is to build trust within communities by appropriately listening and responding to stated values and interests.  Examples could be respecting data sovereignty in perpetuity and providing community-wide STEM educational opportunities.


\section{Approach and Guiding Principles}\label{sec:approach}

The criteria described above are an important component of the broader project to identify and evaluate potential locations for Cosmic Explorer. In this section, we provide a brief description of the approach and principles that will guide that project, which will be carried out in consultation with the National Science Foundation, governments entities, and the local communities and organizations with an investment in the project. The detailed approach will be the subject of a future paper. 


Community partnerships will \emph{be part of} the location identification and evaluation process and, significantly, not something that begins after a small number of physically promising locations have already been identified.  This approach is adopted in order to respectfully work within the socio-cultural context and evolving legal landscape.

The identification of potential \CE\ locations and initial evaluation process has two phases.
The first phase involves a rigorous suitability analysis of the continental United States using publicly available data.
This combines criteria for the physical requirements, such as flatness, seismic noise and other known environmental factors, with characterizations of the social and cultural landscape, such as quality of life factors and land claims.  
These studies will be ongoing and updated throughout the site identification process and iteratively distilled into communication materials that will be used by the \CE\ team and to connect with communities. 

The second phase  
includes on-the-ground work, beginning with introductory trips to regions of interest that were prioritized during the first phase. 
Prior to these visits we will 
identify communities and organizations with an investment in, or who would be impacted by, an observatory in the region of interest.
These could include local and Indigenous communities, academic institutions, national labs, and local and state governments.
In-person visits will serve the purpose of building relationships with these groups and developing a common vision for \CE{}'s stewardship.
Relationship building will necessarily be a long-term, iterative process in order to align mutual goals.

Partnership with Indigenous nations and communities will follow formal protocol and 
consultation\footnote{\lq Consultation' in this context means the a set of legal or established protocols for approaching, presenting, and discussing potential plans that would impact Indigenous communities. Consultation ensures tribal interests are held central throughout the project.  A fundamental tenet for consultation is that tribes have continual power to consent to or veto plans as new information and choices become available.}
in order to build 
a plan for \CE\ that is compatible both with the lifecycle of major research infrastructure and with a community-based vision. 
This may include plans not only for construction and operations but also for decommissioning and divestment of the observatories and return of the land to its natural state. 
Each region of interest will have a unique timeline that primarily depends on how relationship building with communities unfolds.  

Pending community permission we will move to the next phase.
On-location physical assessment will occur at locations where local and Indigenous communities see positive value in the prospect of potentially hosting a \CE\ facility and are open to collaboratively exploring ways in which \CE\ can support and further their interests.
The goal of the on-location assessments is to collect decisive information about the physical and socio-cultural suitability of locations that can only be obtained locally. 
These evaluations will include environmental noise measurements, assessments of the geological, geotechnical, economic, and environmental impact, and legal aspects associated with a given location. 

This approach will require a significant upfront investment in terms of cost, time, and the number of well-trained personnel.
Each location will be rigorously evaluated against each of the criteria described here and a process, such as a rubric, will be developed to rank those locations.
The expected outcome from the complete process is a report, to be delivered to the National Science Foundation, detailing candidate locations that have strengthening relationships and synergies between \CE\ and their local communities, that have undergone assessment, and which may be suitable for hosting a \CE\ observatory.

\section{Conclusions}

\CE, the United States' next-generation gravitational-wave observatory, is expected to play a central role in gravitational-wave and multi-messenger astronomy over the coming decades. The locations that host \CE\ will be both a part of and in some ways determinate of the project's success. This article presented a set of criteria --- derived from scientific performance, cost boundaries and physical access, and social factors --- for identifying and evaluating the suitability of locations for hosting the \CE\ observatories. 

According to the nominal \CE\ timeline~\cite{evans2023cosmicexplorersubmissionnsf}, site selection would happen in the late 2020s ahead of final design and construction in the early 2030s. The process used to evaluate locations according to these criteria  is expected to be carried out in the mid-2020s by a multidisciplinary team in coordination and consultation with local and Indigenous communities. By 2028 the team expects this process will lead to a report on the suitability of locations for hosting \CE\ and their ranking, which will be sent to the Director of the NSF for final selection.  As the process for evaluating locations is further developed and implemented we plan to report on it in more detail. 

\begin{acknowledgments}
The authors are grateful to Alessandra Corsi for her review of a final draft of this article. Thanks to Evan Hall for fruitful discussions on aspects of the \CE\ facilities and Newtonian noise. JRS acknowledges support from the Dan Black Family Trust, Nicholas and Lee Begovich, and NSF Grants No. PHY-2308985 and AST-2219109. 
VM and KP acknowledge support from the NSF Grant PHY-2308988. 
JBR, CS, and SB acknowledge support from NSF Grant PHY-2308989. 
ME, KK, and DS acknowledge support from NSF Grant No. PHY-2309064.
AS, AE, and ML acknowledge support from the NSF Grant PHY -2308990.
KD, WB, JH, CL, and PS acknowledge support from the NSF Grant PHY-2308986. FS research is supported by the NSERC and the FRQNT through the RQMP.
\end{acknowledgments}

\bibliography{references}

\end{document}